\documentclass[reprint,amsmath,amssymbGaps,nofootinbib,onecolumn,notitlepage]{revtex4-1}

\usepackage{amsmath} 
\usepackage{amssymb}
\usepackage{amsfonts}
\usepackage{amstext}
\usepackage{graphicx}
\usepackage{bm}
\usepackage{color} 
\usepackage{transparent}
\usepackage{mathtools}
\usepackage{slashed}
\usepackage{hyperref}

\makeatletter
\renewcommand{\p@subsection}{}
\renewcommand{\p@subsubsection}{}
\makeatother

\numberwithin{equation}{section}

\newcommand{\ze}{\kern 0.05em}

\begin{document}

\title{Charged Black Holes in AdS Spaces in $4D$ Einstein Gauss-Bonnet Gravity}

\author{Pedro G. S. Fernandes}
 \email{p.g.s.fernandes@qmul.ac.uk}
\affiliation{School of Physics and Astronomy, Queen Mary University of London, Mile End Road, London, E1 4NS, UK}


\begin{abstract}

\par Recently a non-trivial 4-dimensional theory of gravity that claims to circumvent Lovelock's theorem and avoid Ostrogradsky instability was formulated in [D. Glavan and C. Lin, Phys. Rev. Lett. 124, 081301 (2020) \cite{Glavan:2019inb}]. This theory, named ``\textit{$4D$ Einstein Gauss-Bonnet gravity}'', presents several novel predictions for cosmology and black hole physics. In this paper, we generalize the vacuum black hole solution of Glavan \& Lin to include electric charge in an anti-de Sitter space and explore some properties of this solution such as the asymptotics, properties of the horizons, the general relativity limit and thermodynamics.

\end{abstract}

\maketitle

%
 \section{Introduction}\label{S1}
%
\par A diversity of observational data is delivering information with unprecedented accuracy on the strong gravity region around black holes (BHs) - see \textit{e.g.} the reviews \cite{Berti:2015itd,Barack:2018yly}. These data include, in particular, the gravitational wave events that have been observed as a result of BH binaries inspiral and merger. Another exciting piece of observational evidence comes from the release of the first image of a BH shadow by the Event Horizon Telescope collaboration \cite{Akiyama:2019eap,Akiyama:2019cqa}. It is then of the uttermost importance to scrutinize good alternatives to the standard General Relativity (GR) paradigm.
\par When approaching the low energy limit, string theories give rise to effective models of gravity in higher dimensions that involve higher order curvature terms in the action. It is speculated that these higher order corrections to Einstein's GR might solve the singularity problem of BHs. Lovelock's theorem \cite{Lovelock} states that if assuming a 4-dimensional spacetime, diffeomorphism invariance, metricity and second-order equations of motion, then GR with a cosmological constant is the unique theory of pure gravity \cite{Glavan:2019inb}. In a recent letter by Glavan \& Lin \cite{Glavan:2019inb}, however, a novel theory of gravity in 4-dimensional spacetime named "4D Einstein Gauss-Bonnet gravity" (4D EGB), obeying all previous restrictions is obtained by circumventing Lovelock's theorem. By rescaling the coupling constant of the Gauss-Bonnet term by a factor of $1/(D-4)$, where $D$ is the number of spacetime dimensions, and defining the four-dimensional theory as the limit $D \to 4$, the Gauss-Bonnet term gives rise to non-trivial dynamics. Remarkably, 4D EGB preserves the number of graviton degrees of freedom, thus being free from Ostrogradsky instability \cite{Glavan:2019inb}. Note that this theory may be considered a particular case of $f(\mathcal{G})$ gravity (with $\mathcal{G}$ the Gauss-Bonnet term) studied, \textit{e.g.}, in \cite{Nojiri:2005jg,Nojiri:2001aj,Cvetic:2001bk}. The 4D Einstein Gauss-Bonnet model does not resort to any kind of matter or non-minimal couplings as extended-Scalar-Tensor-Gauss-Bonnet models do\cite{Silva_2018,Doneva_2018,Antoniou_2018,Cunha:2019dwb}. This new approach to gravitational dynamics presents several novel predictions for cosmology and BH physics. Particularly, in the latter, a new branch of static spherically symmetric BH solutions is obtained which differs from the standard vacuum-GR Schwarzschild BH. Remarkably, this BH solution had been found initially in different contexts, such as gravity with a conformal anomaly \cite{Cai:2009ua,Cai:2014jea} and in gravity with quantum corrections \cite{Cognola:2013fva,Tomozawa:2011gp}. The model considered in \cite{Glavan:2019inb}, however, is a classical modified gravity model, thus on equal footing with general relativity on this regard. Amongst some remarkable properties, these novel BHs are "practically free" from the singularity problem as the gravitational force is repulsive at small distances and thus an infalling particle never reaches the singularity. Studies of the novel theory include \cite{MinyongCheng,Konoplya:2020bxa} which probe the innermost stable circular orbit (ISCO), the BH shadow, its stability and the quasi-normal modes. In $D\geq 5$ spacetime dimensions, BH solutions have been obtained, in the context of Einstein Gauss-Bonnet models, in vacuum \cite{Boulware}, linear \cite{Wiltshire2,Cvetic:2001bk} and non-linear electrodynamics \cite{Wiltshire}, anti-de Sitter (AdS) spaces \cite{Cai:2001dz} and there have been attempts to construct rotating solutions, \textit{e.g.}, \cite{2008PhLB..661..167B,2009CQGra..26f5002A}.
\par The AdS/CFT correspondence \cite{Maldacena:1997re,Gubser:1998bc,Witten:1998qj,Witten:1998zw} remains one of the most important developments in the last decades. Black holes in AdS spaces have attracted a great deal of attention in recent times as it has been argued that thermodynamics of BHs in AdS spaces can be identified with the thermodynamics of a certain CFT at high temperature \cite{Cai:2001dz}. A Schwarzschild BH in an AdS space is thermodynamically unstable when the horizon radius is small, while it is stable for large radius; there is a phase transition, named \textit{Hawking-Page phase transition} \cite{Hawking:1982dh} which is explained by Witten \cite{Witten:1998zw} as the confinement/deconfinment transition of the Yang-Mills theory in the AdS/CFT correspondence. So far, neither charged BHs nor BHs in AdS spaces have been studied in $4D$ Einstein Gauss-Bonnet models, which is the aim of this paper.
\par The paper is organized as follows. In section \ref{S2} we review the novel approach to the Einstein Gauss-Bonnet theory in 4-dimensional spacetime and the static vacuum BH solution obtained in \cite{Glavan:2019inb}. Next we generalize this result to obtain a charged BH solution in AdS space in the $4D$ EGB theory and discuss some of its properties, asymptotics and limits. In section \ref{S3} we briefly discuss some thermodynamical properties for this novel BH, such as the Hawking temperature and entropy. Final remarks are presented in section \ref{Conclusions}.

%
\section{The $4D$ Einstein Gauss-Bonnet model and charged solutions in AdS spaces}\label{S2}
%
\par The Einstein-Gauss-Bonnet theory in $D$ dimensions is described by the action
\begin{equation}
S=\frac{1}{16\pi}\int d^{D} x \sqrt{-g}\left[R+ \alpha \mathcal{G}\right],
\end{equation}
where $\alpha$ is the coupling constant of the Gauss-Bonnet term, defined as
\begin{equation}
\mathcal{G}=R^2 - 4 R_{\mu \nu}R^{\mu \nu} + R_{\mu \nu \rho \sigma}R^{\mu \nu \rho \sigma},
\end{equation}
with $R$ the Ricci scalar, $R_{\mu \nu}$ the Ricci tensor and $R_{\mu \nu \rho \sigma}$ the Riemann tensor and $g$ the determinant of the metric $g_{\mu \nu}$. For $D=4$ dimensions the integral over the Gauss-Bonnet term is a topological invariant, thus not contributing to the dynamics. However, as found recently in \cite{Glavan:2019inb}, by rescaling the coupling constant as
\begin{equation}
\alpha \to \frac{\alpha}{D-4},
\label{eq:rescale}
\end{equation}
and then considering the limit $D\to 4$, Lovelock's theorem is circumvented and new spherically symmetric BH solutions emerge. The static and spherically symmetric BH solution found in \cite{Glavan:2019inb} is of the form
\begin{equation}
d s^{2}=-e^{2A(r)} d t^{2}+e^{2B(r)}dr^2+r^{2} d \Omega_{2},
\end{equation}
with
\begin{equation}
e^{2A(r)}=e^{-2B(r)}=1+\frac{r^2}{2\alpha}\left(1\pm \sqrt{1+\frac{8M\alpha}{r^3}}\right),
\label{eq:solOld}
\end{equation}
with $M$ the BH mass. Note the existence of two branches for the BH solution, depending on the sign chosen for the square-root, thus being labeled as the "positive" or "negative" branches. For negative values of $\alpha$ there is no real solution at short radial distances for which $r^3<-8M\alpha$. Also, in the limit of vanishing coupling constant $\alpha$ and in the far field limit we can only recover the GR Schwarzschild BH for the negative branch. Henceforth we shall only consider $\alpha>0$ and focus mainly on the negative branch solutions.

\subsection{Charged black hole in AdS space in the $4D$ EGB model}
\par Consider now the Einstein-Maxwell Gauss-Bonnet theory in $D$ dimensions with a negative cosmological constant $\Lambda_0=-\frac{(D-1)(D-2)}{2l^2}$ given by the action
\begin{equation}
S=\frac{1}{16\pi}\int d^{D} x \sqrt{-g}\left[R+ \frac{(D-1)(D-2)}{l^2} +\frac{\alpha}{D-4} \mathcal{G} - F_{\mu \nu}F^{\mu \nu}\right],
\label{eq:action}
\end{equation}
where $F_{\mu \nu}=\partial_\mu A_\nu - \partial_\nu A_\mu$ is the usual Maxwell tensor. We consider a static, spherically symmetric metric ansatz in $D$ dimensions given by
\begin{equation}
d s^{2}=-e^{2A(r)} d t^{2}+e^{2B(r)}dr^2+r^{2} d \Omega_{D-2},
\label{eq:ansatz}
\end{equation}
and an electrostatic vector potential $A=V(r) dt$. All functions are radial dependent only and we shall omit this dependence henceforth. Substituting the metric ansatz \eqref{eq:ansatz} and the electrostatic potential in the action \eqref{eq:action} we notice the existence of a first integral
\begin{equation}
V'(r)=-\frac{Q}{r^{D-2}}e^{A+B},
\end{equation}
with $Q$ an integration constant interpreted as the electric charge measured at infinity, and the action reduces to the remarkably simple form
\begin{equation}
S=\frac{\Sigma_{D-2}}{16\pi} \int dt dr\, e^{A+B}(D-2)\left[r^{D-1}\psi\left(1+\alpha (D-3) \psi \right) + \frac{r^{D-1}}{l^2} + \frac{2Q^2 r^{3-D}}{(D-3)(D-2)} \right]^\prime ,
\label{eq:effaction}
\end{equation}
with the prime denoting a radial derivative, $\Sigma_{D-2}=\frac{2\pi^{\frac{D-1}{2}}}{\Gamma \left[ \frac{D-1}{2}\right]}$ the surface area of the $(D-2)$-dimensional hypersurface $d\Omega_{D-2}$ and
\begin{equation}
\psi = r^{-2}\left(1-e^{-2B} \right).
\end{equation}
From the action \eqref{eq:effaction} one can find the solution
\begin{equation}
e^{A+B}=1,
\end{equation}
\begin{equation}
\psi\left(1+\alpha (D-3) \psi \right) + \frac{1}{l^2} + \frac{2Q^2 r^{4-2D}}{(D-3)(D-2)} = \frac{16\pi M}{(D-2)r^{D-1} \Sigma_{D-2}},
\end{equation}
with $M$ the ADM mass. Taking the limit $D \to 4$, we obtain the exact solution in closed form
\begin{equation}
\begin{aligned}
  -g_{00}=e^{2A}&=e^{-2B}=1+\frac{r^2}{2\alpha}\left(1\pm \sqrt{1+4\alpha\left(\frac{2M}{r^3}-\frac{Q^2}{r^4}-\frac{1}{l^2}\right)}\right), \\
  A &= \frac{Q}{r} dt,
\end{aligned}
\label{eq:BHsolution}
\end{equation}
which remarkably coincides with the one obtained in \cite{Cai:2009ua,Cai:2014jea} for gravity with a conformal anomaly and resembles the solution found by Wiltshire \cite{Wiltshire2,Wiltshire} much before. An alternative derivation of the charged Einstein Gauss-Bonnet black hole solution with a cosmological constant in an arbitrary number of dimensions can be found in ref. \cite{Cvetic:2001bk}. The solution \eqref{eq:BHsolution} could be obtained directly from the derivation done in ref. \cite{Cvetic:2001bk} provided that the Gauss-Bonnet coupling constant is rescaled as in eq. \eqref{eq:rescale}, which in the authors' notation amounts to $c \to c/(d-3)$.

\begin{figure}[ht!]
\centering
\includegraphics[width=.5\textwidth]{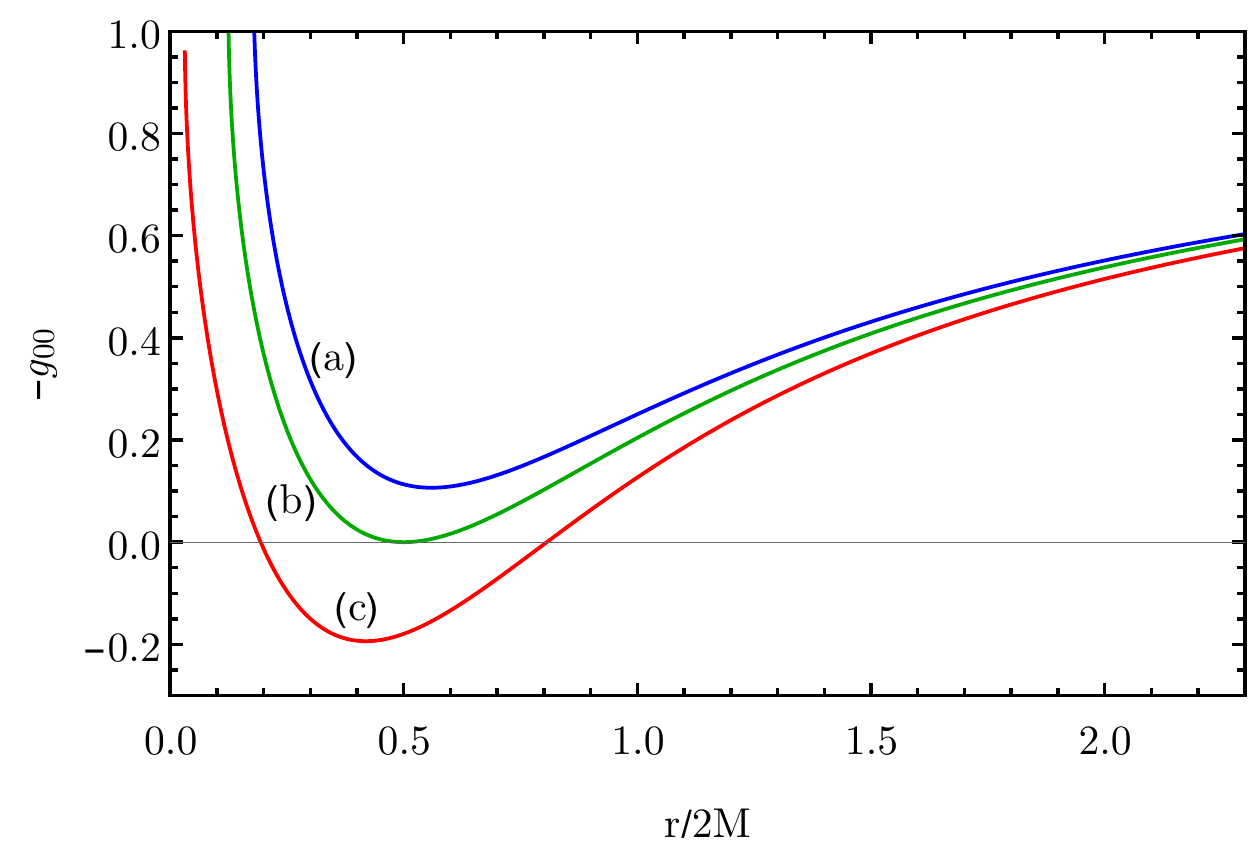}
\caption{Metric function $-g_{00}$ as a function of the radial coordinate for several values of $M_*/M$ for the minus sign branch, in the absence of a cosmological constant. (a, blue) $M<M_*$; (b, green) $M=M_*$; (c, red) $M>M_*$.}
\label{fig:metrics}
\end{figure}

There are two branches of solutions. The one with the plus sign behaves asymptotically as
\begin{equation}
-g_{00} \sim 1 + \frac{2M}{r\sqrt{1-\frac{4\alpha}{l^2}}} - \frac{Q^2}{r^2\sqrt{1-\frac{4\alpha}{l^2}}} + \frac{r^2}{2\alpha}\left(1+\sqrt{1-\frac{4\alpha}{l^2}} \right) + \mathcal{O}\left(\frac{1}{r^3}\right),
\end{equation}
while the solution with the minus sign behaves as
\begin{equation}
-g_{00} \sim 1 - \frac{2M}{r\sqrt{1-\frac{4\alpha}{l^2}}} + \frac{Q^2}{r^2\sqrt{1-\frac{4\alpha}{l^2}}} + \frac{r^2}{2\alpha}\left(1-\sqrt{1-\frac{4\alpha}{l^2}} \right) + \mathcal{O}\left(\frac{1}{r^3}\right).
\end{equation}
In the first case, in the limit of vanishing cosmological constant, we asymptotically obtain a Reissner-Nordstr\"om-AdS solution with negative gravitational mass and imaginary charge, while the latter reduces to the Reissner-Nordstr\"om solution with positive gravitational mass and real charge. In the limit of vanishing mass and charge one obtains
\begin{equation}
-g_{00} \sim 1 + \frac{r^2}{2\alpha}\left(1\pm \sqrt{1-\frac{4\alpha}{l^2}} \right) + \mathcal{O}\left(\frac{1}{r^2}\right),
\end{equation}
from which, assuming a positive value, the coupling constant must obey $0\leq \alpha \leq l^2/4$, beyond which this theory is undefined. Thus, we arrive at two AdS solutions with effective cosmological constants \cite{Cai:2001dz}
\begin{equation}
l_{eff}^2=\frac{l^2}{2}\left(1\pm \sqrt{1-\frac{4\alpha}{l^2}} \right),
\end{equation}
and these coincide in the limit $\alpha=l^2/4$, for which $-g_{00} \sim 1+r^2/l_{eff}^2$. On the other hand, if $\alpha<0$, the solution is still an AdS space if one takes the minus sign, but becomes a de Sitter space if one takes the plus sign. The event horizon of the black holes $r_+$ is larger root of the following equation
\begin{equation}
1-\frac{2M}{r}+\frac{Q^2+\alpha}{r^2}+\frac{r^2}{l^2}=0,
\end{equation}
which in the absence of a cosmological constant, has the simple solution
\begin{equation}
r_{\pm}=M \pm \sqrt{M^2-Q^2-\alpha}.
\label{eq:horizons}
\end{equation}
For a non-vanishing cosmological constant the expression for $r_+$ is complicated a not particularly elucidative, thus we do not present it here. The physical properties of this branch differ depending whether the mass $M$ is larger or smaller than
a critical mass given by
\begin{equation}
M_* = \sqrt{Q^2+\alpha}.
\label{eq:critmass}
\end{equation}
To illustrate, in Fig. \ref{fig:metrics} we plot the radial dependence of the metric function $-g_{00}$ (for the branch with the negative sign) for three situations: one with $M<M_*$, other with $M>M_*$ and the extremal case $M=M_*$.
If $M<M_*$ there are no horizons, and thus no BH solutions, while for $M>M_*$ there are two horizons as given in \eqref{eq:horizons}, and for $M=M_*$ there is one degenerate horizon, corresponding to an extremal BH. In the limit $\alpha \to 0$ we recover, for the negative sign branch, the Reissner-Nordstr\"om-AdS metric from general relativity
\begin{equation}
-\lim_{\alpha\to 0} g_{00} = 1-\frac{2M}{r}+\frac{Q^2}{r^2}+\frac{r^2}{l^2}.
\end{equation}
Particularly important limits of the theory are:
\begin{enumerate}
\item In the case of vanishing charge $Q$ and cosmological constant, we recover \ref{eq:solOld} first obtained in ref. \cite{Glavan:2019inb}.
\item For a vanishing cosmological constant we obtain the electrically charged solution of the theory, analogous to the Reissner-Nordstr\"om BH from GR electro-vacuum.
\item For both vanishing mass and charge we arrive at the AdS solutions of the model.
\item In the limit $\alpha \to 0$, we recover GR solutions.
\end{enumerate}

%
\section{Some thermodynamical properties}\label{S3}
%
\par In this section, we briefly explore the thermodynamics of these novel BHs. Henceforth, we shall restrict our discussion to the \textbf{negative branch} of solution, as one recovers the general relativity solutions in the appropriate limits. BH thermodynamics may provide useful insights into quantum properties of gravitational field, in particular, the thermodynamics of AdS black holes have been of great interest since the pioneering work by Hawking and Page \cite{Hawking:1982dh}, who suggested the existence of a phase transition in AdS black holes. The thermodynamics of charged Einstein Gauss-Bonnet black holes with a cosmological constant are discussed in detail in an arbitrary number of dimensions in ref. \cite{Cvetic:2001bk} and the expressions for the thermodynamic quantities should be valid in the four-dimensional limit of the theory provided that the Gauss-Bonnet coupling constant is rescaled as in eq. \ref{eq:rescale}.

We can express the BH mass $M_+$ in terms of $r_+$ by solving $g_{00}|_{r=r_+}=0$, resulting in
\begin{equation}
M_+=\frac{r_+}{2}\left(1+\frac{r_+^2}{l^2} + \frac{Q^2+\alpha}{r_+^2} \right).
\end{equation}
The BH has a Hawking temperature given by
\begin{equation}
T=\frac{\kappa}{2\pi},
\end{equation}
where $\kappa$ is the surface gravity given by \cite{Kumar:2018vsm} $\kappa^2=-\frac{1}{2}\nabla_\mu \xi_\nu \nabla^\mu \xi^\nu$, with $\xi^\mu$ a killing vector, which for a static, spherically symmetric case takes the form $\xi^\mu=\partial_t^\mu$. For our metric ansatz \eqref{eq:ansatz} the surface gravity is
\begin{equation}
\kappa = \frac{1}{2} \frac{d}{dr} (e^{2A})|_{r=r_+},
\end{equation}
resulting in a Hawking temperature given by
\begin{equation}
T_+=\frac{3r_+^4 + l^2(r_+^2-Q^2-\alpha)}{4\pi l^2 r_+ (r_+^2 + 2\alpha)},
\end{equation}
which in the limit of a vanishing cosmological constant has a real root for $M=M_*$ defined in \ref{eq:critmass}. Thus, BHs whose mass is equal to the critical mass are extremal BHs.
\par In GR the entropy $S$ of a BH obeys the Hawking-Bekenstein formula $S=A/4$, where $A$ is the area of the event horizon of the BH. In general, when considering higher order curvature terms, however, the BH entropy does not satistfy the Hawking-Bekenstein relation. To compute the BH entropy we use the approach of \cite{Cai:1998vy,Cai:2001dz}, which is based on the fact that, as thermodynamic systems, black holes must obey the first law of thermodynamics
\begin{equation}
dM=TdS + \sum_i \mu_i dQ_i,
\label{eq:1stlaw}
\end{equation}
where $\mu_i$ are the chemical potentials corresponding to the conserved charges $Q_i$. Using \eqref{eq:1stlaw} one has
\begin{equation}
S=S_0 + \int \frac{1}{T_+} dM = S_0 + \int_0^{r_+} \frac{1}{T_+} \left( \frac{\partial M_+}{\partial r_+'}\right)_{Q_i} dr_+',
\end{equation}
where $S_0$ is an integration constant, which can be fixed by using the argument that the BH entropy should vanish as the event horizon of the BH disappears \cite{Cai:2001dz}. The BH entropy is then given by
\begin{equation}
S = \pi r_+^2 + 2\pi \alpha \log{r_+^2} + \tilde{S}_0 \equiv \frac{A}{4} + 2\pi \alpha \log{\frac{A}{A_0}},
\end{equation}
with $A_0$ some constant with units of area. This expression coincides with the Hawking-Bekenstein formula plus a logarithmic correction term. According to statistical interpretations of the BH entropy in some quantum theories of gravity such as loop quantum gravity and string theory, it can be argued that the leading term of statistical degrees of freedom yield the Hawking-Bekenstein area term, whereas the subleading term is a logarithmic term \cite{Cai:2009ua,Cai:2014jea,Cognola:2013fva}. However, it is quite difficult to produce such a logarithmic term in the BH entropy in some effective local theory of gravity even with higher derivative curvature terms. This theory then provides a possible interpretation for the appearance of a logarithmic term in the entropy. Remarkably, all explicit contributions from the charge $Q$ and cosmological constant cancel out, resulting in a simple expression for the BH entropy $S$. As a remark, the branch with a positive sign, however, does not obey the Hawking-Bekenstein formula plus a simple logarithmic correction for the entropy, presenting a complicated lengthy expression which is not particularly elucidative.
%
\section{Conclusions}\label{Conclusions}
%
In this paper, based on the work of Glavan \& Lin \cite{Glavan:2019inb} on the novel $4D$ Einstein Gauss-Bonnet gravity, we generalized the vacuum BH solution of the model to include electric charge in AdS spaces. This BH solution was derived in analytical closed form. Next we studied the asymptotics of the solution as well as its dependence on the parameters of the model. We found that, for the negative branch, the solution resembles the Reissner-Nordstr\"om BH in the far field in the absence of a cosmological constant and that the model allows solutions with one, two or no horizons depending if the mass is equal, above or below a certain critical mass, respectively. Also, in the limit of vanishing coupling constant $\alpha$ it was found that one recovers the GR Reissner-Nordstr\"om-AdS solution. We studied briefly the thermodynamics of this new BH, obtaining its Hawking temperature and entropy. Remarkably, for the negative branch, the entropy of the solution obeys the Hawking-Bekenstein area formula plus a logarithmic correction term, having the same form as predicted by some quantum gravity theories such as string theory and loop quantum gravity. As an avenue of further research one may further study the thermodynamics of this novel solution as well as the quasinormal modes, shadows and the ISCO as done in \cite{Konoplya:2020bxa,MinyongCheng} for the vacuum solution. As a final remark we would like to point out that to obtain the counterpart BH solutions in de Sitter space one could take $l^2 \to -l^2$.

\medskip
%
\section*{Acknowledgements}
%
\par P.F. is supported by the Royal Society grant RGF/EA/180022 and acknowledges support from  the  project CERN/FIS-PAR/0027/2019. The author would like to thank Carlos A. R. Herdeiro, Eugen Radu, Pedro M. G. Carrilho and David J. Mulryne for useful comments on the manuscript and their support.

\bibliography{biblio}

\end{document}